\documentclass[conference]{IEEEtran}
\ifCLASSOPTIONcompsoc
  \usepackage[nocompress]{cite}
\else
  \usepackage{cite}
\fi
\ifCLASSINFOpdf
   \usepackage[pdftex]{graphicx}
   \graphicspath{{. . /pdf/}{. . /jpeg/}}
   \DeclareGraphicsExtensions{. pdf,. jpeg,. png}
\else
  \usepackage[dvips]{graphicx}
  \graphicspath{{. . /eps/}}
  \DeclareGraphicsExtensions{. eps}
\fi
\usepackage{amsmath}
\usepackage{epsfig}
\usepackage{graphicx}
\usepackage{caption,subfigure}
\usepackage{algorithmic}
\ifCLASSOPTIONcompsoc
  \usepackage[caption=false,font=footnotesize,labelfont=sf,textfont=sf]{subfig}
\else
  \usepackage[caption=false,font=footnotesize]{subfig}
\fi

\usepackage{stfloats}
\begin{document}
%
\title{Family Shopping Recommendation System \\Using User Profile and Behavior Data}

\author{
	\IEEEauthorblockN{Jiacheng Xu}
	\IEEEauthorblockA{ School of Computer Science and\\Software Engineering\\
		East China Normal University\\
		Email: xujiacheng28@outlook.com
	}
}


%


\maketitle

\begin{abstract}
With the arrival of the big data era, recommendation system has been a hot technology for enterprises to streamline their sales. Recommendation algorithms for individual users have been extensively studied over the past decade. Most existing recommendation systems also focus on individual user recommendations, however in many daily activities, items are recommended to the groups not one person. As an effective means to solve the problem of group recommendation problem, we extend the single user recommendation to group recommendation. Specifically we propose a novel approach for family-based shopping recommendation system. We use the dataset from the real shopping mall consisting of shopping records table, client-profile table and family relationship table. Our algorithm integrates user behavior similarity and user profile similarity to build the user based collaborative filtering model. We evaluate our approach on a real-world shopping mall dataset.
\end{abstract}

%
\IEEEpeerreviewmaketitle

\section{Introduction}\label{sec:intro}
With the arrival of the big data era, recommendation system \cite{Su2009A,resnick1997recommender} is a kind of big data technology for enterprises to gain more profit. Recently, recommendation systems are becoming one of the hotspots in the field of recommendation systems.

In recent years, with the recommendation system research and developments  very rapid, there are  more and more various types of recommendation systems such as mobile recommendation systems, news recommendation system, music recommendation system and so on. However, these recommendation systems are only for individual users.  In fact, many daily activities are make up of many users in groups (e.g. watching movies or TV shows, going to a restaurant, vacationing etc.).  Therefore, the recommendation system needs to consider the preferences of each user in the group for recommendation. This kind of system is called group recommendation system \cite{boratto2016group}. It extends the recommendation object from a single user to a group, posing new challenges to recommendation systems. The interests of the group members can be diverse in reality. How to get the common preference of the group members, alleviating the preference conflict among the group members, and make the recommendation result meet the needs of all group members as much as possible, are the key problems \cite{masthoff2011group} to be solved in the group recommendation system. But at present, the research of group recommendation system has received more and more attention.

Nowadays, group recommendation system is more and more common such as Let's Browse (news group recommendation system) \cite{lieberman1999let}, Masthoff's Group Recommender (music group recommendation system) \cite{masthoff2006pursuit}, Intrigue (tourism group recommendation system) \cite{ardissono2003intrigue}, Where2Eat (Food group recommendation system) \cite{guzzi2011interactive} and so on.  From the above example, group recommendation system can be seen everywhere in our lives.

In this paper, we proposed a novel family recommendation system algorithm. We use the dataset from the real shopping mall consisting of shopping records table, client-profile table and family relationship table. Its aim is to bring huge profits to the shopping mall. In fact, the entire data set acquisition process is very difficult. Because when users go shopping in the supermarket, they do not leave their family information. In other words, they do not tell us who is his wife or who is his children. For solving the problem, we determined their family relationship by telephone in the process of collecting the data. Preprocessing is performed to do data cleaning and curation, which leads to the systematic experiments conducted in the paper.

In a nutshell, the main contributions of this work are:

1) To our best knowledge, this is the first work on family level personalized recommendation system, which is useful for the case when a family share a common account or family members will purchase items for the other members in the family.

2) We propose a novel approach for integrating the information from user/family purchase data, activity data and shopping mall event data into a unified model for the family based recommendation problem.

3) We verify our approach on via an empirical study on a real-world dataset and showcase the efficacy of our method.

In Section \ref{sec:intro}, we present the background of recommendation system and group recommendation system and propose a new group recommendation algorithm. The rest of paper is structured as follows: Section \ref{sec:related} presents the commonly used algorithms such as content-based collaborative filtering \cite{balabanovic1997fab} and group-based collaborative filtering.  In section \ref{sec:method}, we specifically explain the detail of our group recommendation system architecture. Section \ref{sec:exp} shows the experimental result by peer algorithms. Finally Section \ref{sec:con} concludes this paper.

\section{Related Work}\label{sec:related}
\subsection{Memory-based Collaborative Filtering}
Memory-based collaborative filtering (CF) algorithms are popular techniques for recommender systems. It involves the user-item database to predict which items users really like.  According to every user who belongs to a group of people with similarity interests, we can find a group of people who share similar interests. By doing so, we can recommend the item to a user by it. In fact, it is neighborhood-based recommendation system \cite{bell2007improved} and very prevalent.

The neighborhood-based CF algorithm uses the following steps. Firstly, calculate the similarity represented by $W_{i,j}$ which reflects correlation or distance between two uses or two items ,$i$ and $j$. Secondly, predict the unknown ratings  of the user or item on a certain item or user by aggregating the neighbors who are $k$ most similar to get the top-N most frequent items as the recommendation, when the task is to generate a top-N recommendation result.
\subsubsection{Similarity Computation}\label{IIA1}
Similarity computation is one of the most important steps in the process of the recommendation system. Different methods that get the similarity lead to different results. For a user-based collaborate filtering algorithm, we calculate the similarity, $W_{u,v}$, between the users $u$ and $v$ who have both rated the same items. While for an item-based collaborate filtering algorithm \cite{sarwar2001item}, we calculate the similarity, $W_{i,j}$, between the items $i$ and $j$ who have both been rated by the same users. Now, correlation-based similarity and vector cosine-based similarity are fluent and efficient methods.

\begin{figure}[!tb]
	\centering
	\includegraphics[width=0.68\linewidth]{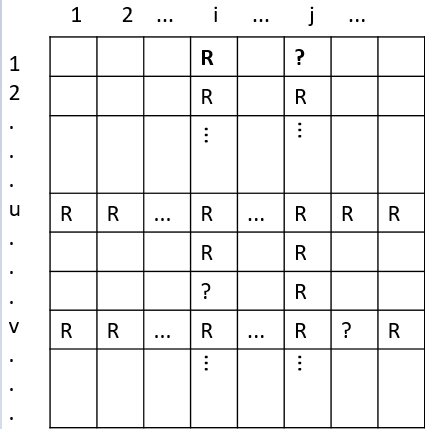}
	\caption{Users-items rating matrix.}
	\label{users-items}
\end{figure}

\textbf{Vector cosine-based similarity} is a useful method to calculate the similarity as popular as the correlation-based similarity we discuss below. The similarity between two items or users can be calculated by treating each items or users as a vector  and computing the cosine of the angle between two vectors\cite{salton1986introduction}. Hence the element inside the vector of the item is the user's rating of the item. Vector cosine similarity between items $i$ and $j$ is given by
\begin{equation*}
\begin{aligned}
W_{i,j}=\cos(\vec{i},\vec{j})=\frac{\vec{i}\cdot \vec{j}}{\left \| \vec{i} \right \|*\left \| \vec{j} \right \|}\\
=\frac{\sum_{u\in U}r_{u,i}r_{u,j}}{\sqrt{\sum_{u\in U}r_{u,i}^2}\sqrt{\sum_{u\in U}r_{u,j}^2}}
\end{aligned}
\end{equation*}
where $\cdot$ denotes the dot-product of the two item vectors $i$, $j$ and $U$ is the whole set of users and $\bar{r_{u,i}}$ is the rating that user $u$ give the item $i$, which is shown in Fig.\ref{users-items}.

But the vector cosine-based similarity algorithm has some limitations. In practice, different users may rate some items using different rating standard. As a result, the vector cosine similarity may not represent the similarity between two items. Taking this into account, we adjust the vector cosine-based similarity algorithm to address this shortcoming. Simply, we can subtract the corresponding user average from each co-rated pair. In fact, the adjusted vector cosine similarity has the same formula as the Pearson correlation similarity. In the following, correlation-based similarity will be introduced in detail.

\textbf{Correlation-based similarity} is the method that we calculate the similarity by Pearson correlation. As we known, Pearson correlation denotes that  two variables linearly relate with each other. For the user-based algorithm, the Pearson correlation between users $u$ and $v$:
\begin{equation*}
\begin{aligned}
W_{u,v}=\frac{\sum_{i\in I}(r_{u,i}-\bar{r_{u}})(r_{v,i}-\bar{r_{v}})}{\sqrt{\sum_{i\in I}(r_{u,i}-\bar{r_{u}})^2}\sqrt{\sum_{i\in I}(r_{v,i}-\bar{r_{v}})^2}}
\end{aligned}
\end{equation*}
where $I$ is the whole set of items, $\bar{r_{u}}$ is the average rating of the items of user $u$ and $r_{u,i}$ is the rating that use $u$ give to  the item $i$.  It show in Fig.\ref{users-items}.

For the item-based algorithm, the Pearson correlation between users $u$ and $v$:
\begin{equation*}
W_{i,j}=\frac{\sum_{u\in U}(r_{u,i}-\bar{r_{i}})(r_{u,j}-\bar{r_{j}})}{\sqrt{\sum_{u\in U}(r_{u,i}-\bar{r_{u}})^2}\sqrt{\sum_{u\in U}(r_{u,j}-\bar{r_{u}})^2}}
\end{equation*}
where $U$ is the whole set of users and $\bar{r_{i}}$ is the average rating of the items $i$. It show in Fig.\ref{users-items}.

In a word, the Pearson correlation-based CF algorithm as a typical collaborate filtering algorithm is so efficient that it is widely used in the Recommendation system \cite{resnick1994grouplens, mclaughlin2004collaborative}.

But the above two methods need the rating matrix about users and items. In a real case, the rating data is not easily available and often is very sparse. However, we can still obtain useful information from implicit data. For example, if two users buy the same goods, they have some common preferences in a shopping mall. The next method  can address such a problem by calculating the precise similarity.

\textbf{Jaccard similarity} \cite{seifoddini1991production} is very simple algorithm. Its main idea is that the more same goods purchased by the two users and the fewer amount of goods bought by the two users, the higher the similarity between them.  It is very useful to process the shopping records without ratings. Jaccard similarity formula between items $i$ and $j$ is given by:
\begin{equation*}
W_{u,v}=\frac{\left | N_{u}\bigcap N_{v} \right |}{\left | N_{u} \bigcup N_{v}  \right |}
\end{equation*}
where $N_{u}$ is the set of items which user $u$ purchase and $N_{v}$ is the set of items which user $v$ purchase.  $\left | N_{u}\bigcap N_{v} \right |$  denotes the same items which are bought by users $u$, $v$. And $\left | N_{u} \bigcup N_{v}  \right |$ means that the amount items which are bought by users $u$, $v$.  If users $u$, $v$ do not buy the same items, their similarity is 0. If users $u$, $v$ do not buy the different items, their similarity is 1.  So we find it can reflects the similarity between users $u$, $v$ effectively.

\textbf{Euclidean distance similarity}\cite{fouss2007random} is another similarity algorithm. It does not need the relationship between users and items.  Each item or user has its own attributes. Therefore we can calculate the user similarity by users' profile by Euclidean distance. Each user attribute can be composed of a vector.  For example, the vector of use $u$ can be denoted by $\vec{u}=(a_1, a_2,. . . , a_n)$. In brief, the similarity problem is transformed into two vector distance problems. Then we decide to  choose the Euclidean distance.  But what we need to pay attention to is that the more distance, the less similarity.

\subsubsection{Prediction Computation}
Prediction computation is also one of critical parts in a collaborative filtering system \cite{herlocker1999algorithmic}. In fact, many methods share similar ideas. In the neighborhood-based CF algorithms, one can choose a subset of the nearest neighbors of the user according to their similarity with another one and aggregate their ratings to generate predictions for user. We can take a weighted average of all the ratings on that item $i$ to make a precise prediction for the rating of the user $k$, on a certain item $i$. The following formula is given by \cite{resnick1994grouplens}:
\begin{equation*}
r_{k,i}=\bar R_{k} + \frac{\sum_{u \in U}(r_{u,i}-\bar r_{u})\cdot w_{k,u}}{\sum_{u \in U}\left | w_{k,u} \right |}
\end{equation*}
where $w_{k,u}$ is the weight between the user $k$ and user $u$, $\bar r_{k}$ is the average ratings for the user $k$ on all other rated items as same as $\bar r_{u}$.
$U$ is the set of users who have rated the item $i$. For memory-based prediction,
we can also  predict the rating simply by the following formula:
\begin{equation*}
r_{k,i}=\frac{\sum_{u \in U}r_{u,i}\cdot w_{k,u}}{\sum_{u \in U}\left | w_{k,u} \right |}
\end{equation*}
\subsubsection{Top-N Recommendations}
The Top-N Recommendations are that a set of N top-ranked  items, which attract the specific  users, are recommended. For example, when you use NetEase cloud music, if you are an old user, you will be recommended a list of songs which you may be interested in.  The Top-N Recommendations can mainly divided into two classes, User-Based Top-N Recommendation Algorithms and Item-Based Top-N Recommendation Algorithms.

For the former, we first choose the $k$ most similar users to the specific user by using the \textbf{Pearson Correlation Similarity} or \textbf{Vector Cosine-based Similarity} \cite{sarwar2000analysis}. After the $k$ most similar users have been chosen, we can aggregate their corresponding rows in the user-item rating matrix $R$ to identify a set of items, $A$. With the set $A$, user-based collaborate filtering techniques recommend the items with the top-N highest scores in $A$ which the specific user has not purchased.

For the latter, we firstly choose the $k$ most similar items for each item according to their similarities predicted by \textbf{Pearson Correlation Similarity} or \textbf{Vector Cosine-based Similarity}. Then we can identify the set, $A$, by taking the union of the k most similar items and removing each of the items in the set, $U$, which the specific user has already purchased. The set $A$ is candidates of recommended items. Then we predict the similarities between each item of the set $A$ and the set $U$. We can recommend the Top-N items \cite{karypis2001evaluation} according to the resulting set of the items which are sorted in decreasing order of the ratings in $A$. The above is the summary concept of the Top-N Recommendations.
\subsection{Group-based Collaborative Filtering}
Group-based Collaborative Filtering is the important point in this paper. It is different from the single user recommendation system, especially the calculating similarity method. Sometimes, it can solve the Cold-Start problem \cite{Miao2016Joint}. Group detection and group preference aggregation are two important part of the Group-based Collaborative Filtering.
\subsubsection{Group Detection}
In most of the existing group recommendation systems, it has identified the grouping \cite{o2001polylens,freyne2006cooperating}. However, in many group recommendation systems, the grouping is not given. So  we should add a user to a certain group by some method. We can call it group detection. There are two commonly used methods: the first is based on user preferences and the second is based on demographic methods. In fact, it is the cluster problem. In this field, the K-means is the most famous method. For example, in \cite{boratto2010groups}, the user-item scoring matrix is directly input to the K-means clustering algorithm, and the users are clustered according to user preferences. And demographic information can also be the input to the K-means clustering algorithm. For example, the group recommendation system Intrigue \cite{ardissono2003intrigue} divides the group into different subgroups  according to the demographic characteristics of the user. But in my experience, the grouping is given  by investigating families.

\subsubsection{Group Preference Aggregation Strategy}
Group preference aggregation which is proposed by Dyer et al. \cite{dyer1979group} is the most important part of the group recommendation system. In the group recommendation systems, preference aggregation refers to the extraction of group preference according to the preferences of group members \cite{dyer1979group}. In an actual world, group recommendation systems need to meet the different requirements by the different group preference aggregation strategies. Meanwhile, it also need to meet the overall satisfaction, fairness, understandability and other requirements.

Recently, some basic group preference aggregation strategies in group recommendation systems have been studied \cite{baltrunas2010group}. In \cite{masthoff2004group}, ten group preference aggregation strategies are analyzed in detail. Then five commonly used strategies are discussed below.

\textbf{Average Strategy:} The average of the group membership ratings is taken as the rating of this group.

\textbf{Most Pleasure Strategy:} The maximum of the group membership ratings is taken as the rating of this group.

\textbf{Least Misery Strategy:} The minimum of the group membership ratings is taken as the rating of this group \cite{berry2010netflix}.

\textbf{Average Without Misery Strategy:} The average of the group membership ratings which is higher than the certain threshold is taken as the rating of this group\cite{mccarthy1998musicfx}.

\textbf{Most Respected Person Strategy:} The rating of group is decided by the ratings of the most respected person in this group.

\subsubsection{Group Preference Aggregation Method}
Group Preference Aggregation which occurs at different stage of the recommendation  is also different. Paper \cite{jameson2007recommendation} summarizes Group Preference Aggregation Method into three categories: recommendation result aggregation, rating aggregation and group preference modeling.

For recommendation result aggregation and rating aggregation, we first generate the recommendation for each user of the group or calculates the predictive ratings, and then generate a group recommendation aggregation or group prediction ratings.

For the group preference model, it first generates a group preference model by aggregating the group members preferences models.  Then we can recommend the items by the group preference model.

In this paper, we mainly aggregate the attributes of the numbers of the family group by a certain strategy and then recommend the items to the family group.
\section{Proposed Approach}\label{sec:method}
\subsection{Datasets}\label{subsec:dataset}
The dataset consist of client profile, transaction dataset, visit dataset, family dataset and participation dataset from the real shopping mall.

\textbf{Client profile dataset} has 4505 client profiles with seven attributes: join days, sex, age, phone, email, neighborhood, register source and income. Its main contents are the client profile  which are used to calculate the user similarity.

\textbf{Transaction dataset} contains 25550 shopping records with categories:
\emph{productbrand}, \emph{producttype} and \emph{maincategory}.The items distribution with three categories show in the Table \ref{brand}, \ref{type}, \ref{category}. It is used to calculate the user similarity by Jaccard similarity algorithm.

\textbf{Visit dataset} has 60427 records with check-in timestamp and checkout timestamp.

\textbf{Participation dataset} records the activities each user participates.

\textbf{Family dataset} has 6378 family groups.  Every family group data contains the information about the composition of the family.

\begin{table}[!tb]
\caption{Item Distribution-Product Brand}
\label{brand}
\centering
\begin{tabular}{|c|c|}
\hline
Product Brand & Amount\\
\hline
DXNIKE &  2435\\
\hline
DXDISNEY & 2003\\
\hline
JEEP SPIRIT & 1629\\
\hline
G\&H & 1578\\
\hline
K\&C & 1437\\
\hline
BAB & 	1258\\
\hline
DBN &	1138\\
\hline
NEW BALANCE &	1005\\
\hline
L &	939\\
\hline
... & ...\\
\hline
\end{tabular}
\end{table}

\begin{table}[!t]
\caption{Item Distribution-Product Type.}
\label{type}
	\centering
\begin{tabular}{|c|c|}
	\hline	
	Product Type &	Amount\\
\hline	
Miscellaneous &	6108\\
\hline
CLOTHING &	5715\\
\hline
INFANT \& KIDS &	3838\\
\hline
SHOES &	2808\\
\hline
SPORTSWEAR &	1404\\
\hline
ACCESSORIES &	1314\\
\hline
TOYS &	1198\\
\hline
FOOD &	664\\
\hline
BAB TOY &	643\\
\hline
CASUALWEAR &	609\\	
\hline
... & ...\\
\hline
\end{tabular}
\end{table}

\begin{table}[!t]
\caption{Item Distribution-Main Category.}
\label{category}
   \centering
	\begin{tabular}{|c|c|}
\hline
Main Category&	Amount\\
\hline
INFANT \& KIDS &	8030\\
\hline
LIFESTYLE &	4021\\
\hline
MENS WEAR &	2861\\\hline
Miscellaneous &	2301\\\hline
KIDS &	1805\\\hline
HOME \& LIFESTYLE &	1332\\\hline
TOYS &	766\\\hline
KNITWEAR &	620\\\hline
ACCESSORIES &	448\\\hline
TC WOVEN SHIRTS &	394	\\	
\hline
... & ...\\
\hline
	\end{tabular}
\end{table}

\subsection{Data Preprocessing}\label{subsec:datapreprocess}
Data Preprocessing is the important stage in the family recommendation system.
It can not only  help us with numerical calculations but also make the experiment results better.  Data Preprocessing can be divided into two parts.

\subsubsection{Data Cleaning}
Some information of the items with the missing values are deleted or replaced by the certain value such as zero, the average value and so on. It can make our dataset  more complete.

\subsubsection{Data Normalization}\label{IIIB2}
The form of the elements of the dataset is various, so it's hard to deal directly with them.

For the user-behavior feature, we can normalize the data items of the participation dataset and the transaction dataset to triples (\emph{memberid}, \emph{productbrand}\verb|\|\emph{producttype}\verb|\|\emph{maincategory}\verb|\|\emph{activity}
, the quantity of the items).

For the client profile dataset, we can separate the attributes of client into numeric variable and category variable. The category variables such as sex, neighborhood, register source and so on can be encoded by the one-shot encoding. For example, the attribute \emph{sex} with three values can be encoded to the vector( Female: $(0,1)$  Male: $(1,0)$ ) and the  numeric variables can be mapped to 0 to 1 by the normalization method.  For the attribute \emph{telephone} and \emph{email}, we can encode them to 0 or 1 according to whether user left the information.
\subsection{User Behavior-based Similarity Matrix}
According to the user-behavior feature with four categories (\emph{productbrand}, \emph{producttype}, \emph{maincategory}, \emph{activity}), we use the triples which we propose in the section \ref{subsec:datapreprocess}. In order to get all the set of items which the user have bought  as the inputs to the Jaccard similarity algorithm proposed in the section \ref{IIA1}, then can get four user similarity matrix\cite{Li2016Low}, which are respectively denoted as $W_{brand}$, $W_{type}$, $W_{category}$, $W_{activity}$.
\subsection{User Profile-based Similarity Matrix}
In the client profile dataset, using the user profile, we can calculate the user similarity matrix\cite{2016MPMA,2017GLOMA}. Firstly, we can generate the user profile vectors by the method introduced in the section \ref{subsec:datapreprocess}. For example, we can formula the user attributes 
to the vector $\vec{a}=(1, 1, 0, 0, 1, 0, 0, 0, 0, 2, 2, 1. 907638, 0. 567901, 0. 468739, ...)$.  Then we can use the user profile vector to calculate the user distance matrix by the Euclidean distance proposed in the section \ref{subsec:datapreprocess}. Then we normalize all the elements of the user distance matrix to interval$(0,1)$, denoted as $D$.  Finally, we get the user similarity matrix by the formula:
\begin{equation*}
W_{uv}=1-D_{uv} \quad s. t.  \quad u, v\in  U
\end{equation*}
where the dimensions of the matrix $W$ and matrix $B$ are the same and $W_{uv}$ is the similarity between user $u$ and user $v$ which belong to the set of users denoted as $U$. To distinguish the similarity matrix above, we denote the similarity matrix as $W_{profile}$.

\subsection{Hybrid Similarity Matrix Algorithm}
The five similarity matrix ($W_{profile}$, $W_{brand}$, $W_{type}$, $W_{category}$, $W_{activity}$) has already gotten.  Then we can aggregate them to get more better similarity matrix. Because aggregating them which are gotten from different categories can  fully represent the similarity between users. The empirical experimental results suggest the usefulness of our model.

\subsection{Family Recommendation Algorithm}
The single user similarity matrix algorithms have been discussed above. For the group similarity matrix, the group discovering  can be stepped, because we have known the member of the family according to the \textbf{Family dataset}.

Then for the family behavior-based similarity matrix, we combine the triples mentioned in the section \ref{IIIB2} with family units. For example, user $i$ triple is $(i,1,1)$ and user $j$ triple is $(j,4,1)$. The members of the family $f$ are user $i$ and user $j$, so the family $f$ triples are $(f,1,1)$ and $(f,4,1)$. Then we use the triples to get all the set of items  the family have bought  as the inputs to the Jaccard similarity algorithm proposed in the section \ref{IIA1}. Finally we can get four user similarity matrix, which are respectively denoted as $W_{brand}^{F}$, $W_{type}^{F}$, $W_{category}^{F}$, $W_{activity}^{F}$.

\begin{figure}[!tb]
	\centering
	\includegraphics[width=\linewidth]{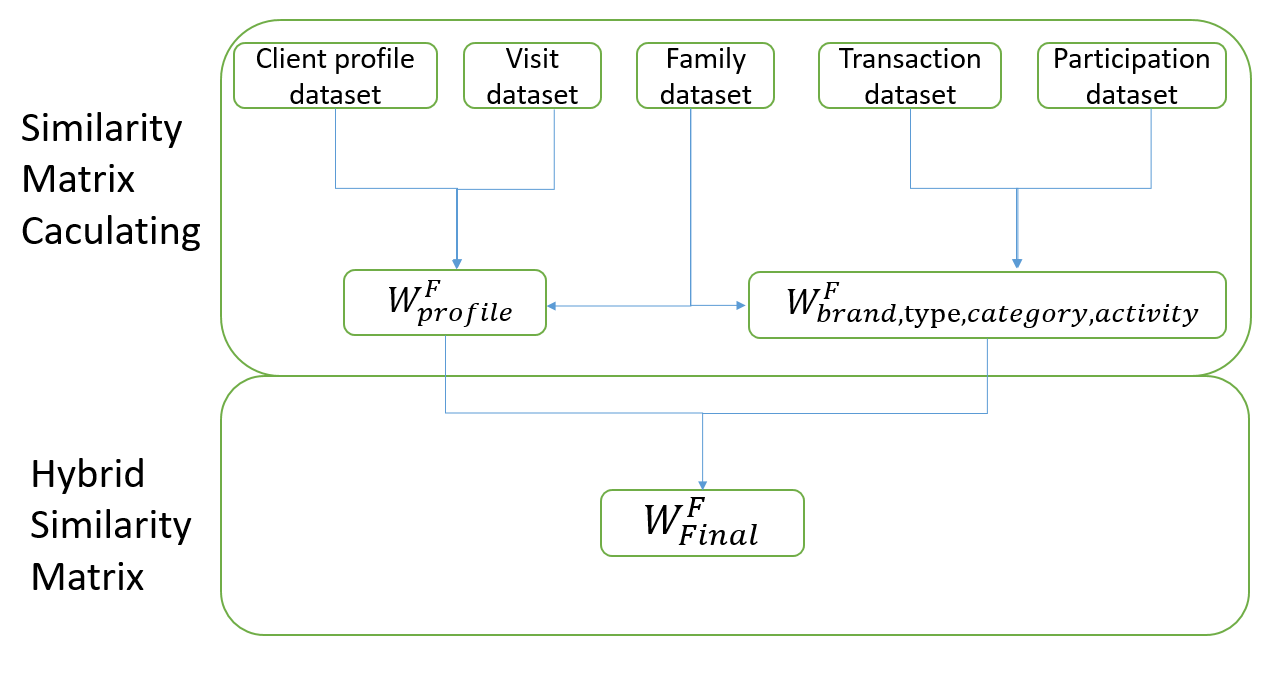}
	\caption{Overview of the family recommendation approach.}
	\label{flowchart}
\end{figure}

For the family profile-based similarity matrix, we firstly sum the vectors of all the member of the family $f$ as the profile vector of the family $f$. We can calculate the similarity matrix like the user profile-based similarity matrix, denoted as $W_{profile}^{F}$. It can make the dissimilar vectors farther apart and  the similar vector more closer.
Finally, we aggregate the five similarity matrix to get the final similarity matrix denoted as $W_{Final}^{F}$ and recommend the item with the family unit using $W_{Final}^{F}$. The whole procedure is shown in the Fig.\ref{flowchart}.

\section{Experiments}\label{sec:exp}
\subsection{Models}
We mainly propose three different models and compare them by the evaluation metric. Finally, we can find that the third model is better than the first one and the second one according to experiment results.

\textbf{User Model:}

User Behavior-based + User Profile-based Recommendation Method

\textbf{Hybrid User Model:}

User Behavior-based+User Profile-base + Hybrid Similarity Matrix Algorithm Recommendation Method

\textbf{Hybrid Family Model:}

User Behavior-based+User Profile-based + Hybrid Similarity Matrix Algorithm + Family Recommendation Method

The recommended items in this experiment have three categories (\emph{productbrand}, \emph{producttype}, \emph{maincategory}). (The details are told in Section \ref{subsec:expp}. \emph{Experimental Procedure}.)

As a result, for the \emph{User Model}, we recommend the item \emph{productbrand} by aggregating the  $W_{brand}$, $W_{activity}$ and $W_{profile}$,recommend the item \emph{producttype} by aggregating the  $W_{type}$, $W_{activity}$ and $W_{profile}$ and recommend the item \emph{maincategory} by aggregating the  $W_{category}$, $W_{activity}$ and $W_{profile}$.

The \emph{Hybrid User Model} and \emph{Hybrid Family Model} recommend the item with three categories by using one similarity matrix. For the \emph{Hybrid User Model}, we recommend the item by aggregating the $W_{brand}$, $W_{type}$, $W_{category}$, $W_{activity}$ and $W_{profile}$. For the \emph{Hybrid Family Model},we recommend the item by aggregating the $W_{brand}^{F}$, $W_{type}^{F}$, $W_{category}^{F}$, $W_{activity}^{F}$ and $W_{profile}^{F}$.

\subsection{Dataset Splitting}
\begin{figure*}[!tb]
	\centering
	\includegraphics[width=0.9\linewidth]{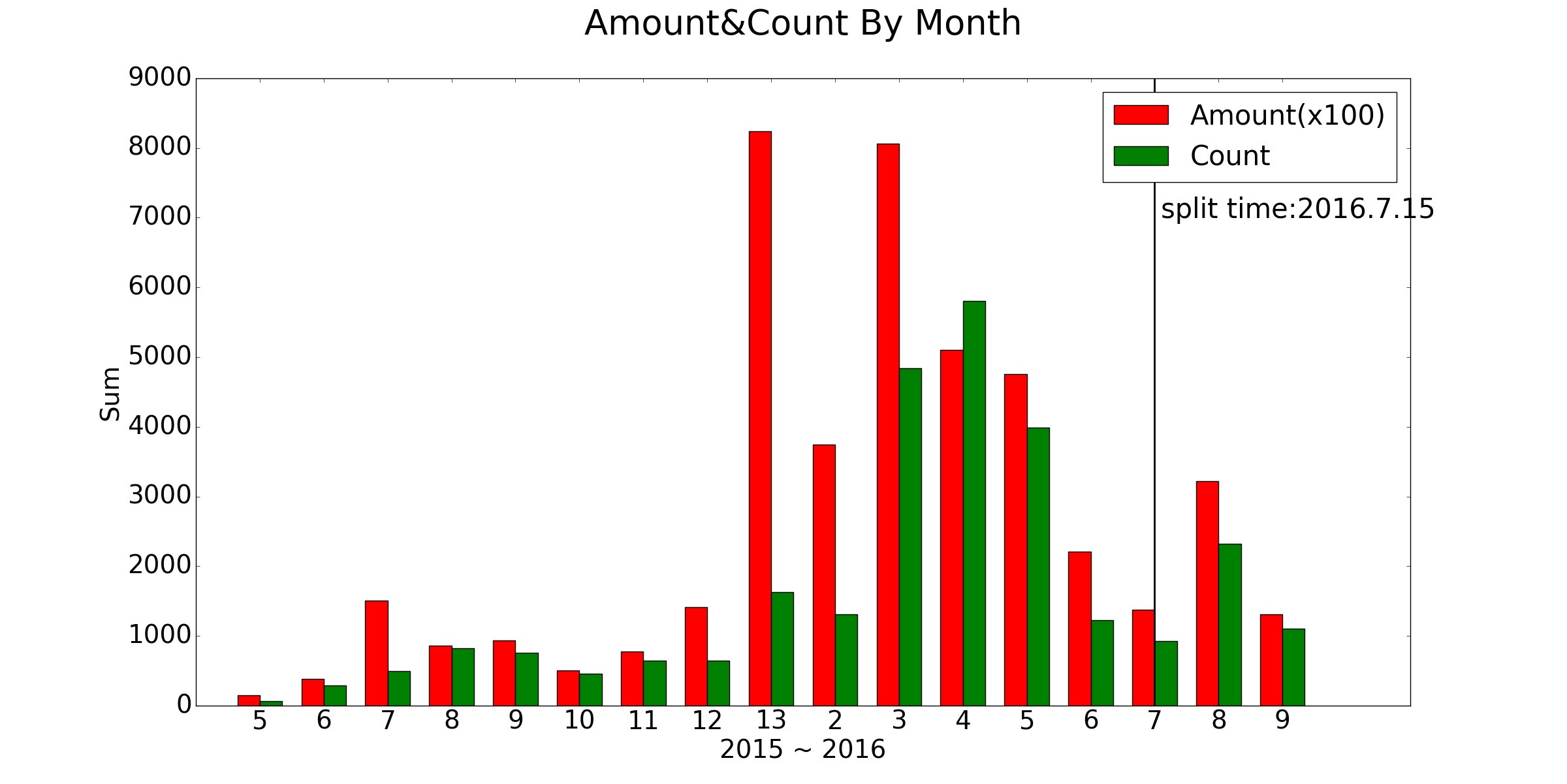}
	\caption{Time and space distribution of the data.}
	\label{datadistribution}
\end{figure*}

Dataset splitting is the important stage when we split the transaction dataset into two parts, one part as testing set and the other part as training set. Because it is the shopping records with timestamps, we decide to split the dataset by the time point. By observing the time and space distribution of the data which is shown in Fig.\ref{datadistribution}., we choose the timestamp `2016-07-15 00:00:00' as the splitting point of the testing set and the training set. Testing set and training set account for 20 percent and 80 percent respectively. Because it is shopping data, testing set and training set split by the time point will be better for us to evaluate the model performances according to the data characteristics.

\subsection{Evaluation Metric}
There are many evaluation metric for recommender system \cite{basu1998recommendation}, such as RMSE (Root Mean Square Error), MAE (Mean Absolute Error), recall rate, precision rate and so on. Here, we finally choose the recall rate and the precise rate for their popularity in related works.

The recall rate describes how many percentage of the user's shopping records are included in the final recommendation list, and the precise rate describes how many  the final recommended lists are user shopping records that have occurred.

\textbf{Recall Rate} is given by
\begin{equation*}
Recall = \frac{\sum_{u \in U}\left|R(u)\bigcap T(u) \right| }{\sum_{u \in U}\left | T(u) \right|}
\end{equation*}

\textbf{Precision Rate} is given by
\begin{equation*}
Recall = \frac{\sum_{u \in U}\left|R(u)\bigcap T(u) \right| }{\sum_{u \in U}\left | R(u) \right|}
\end{equation*}
where $R(u)$ is the set of the items which are recommended to the user $u$ and $T(u)$ is the set of the items which the user $u$ have already bought.
\subsection{Experimental Procedure}\label{subsec:expp}
Our experiment is to recommend the items from three categories (\emph{productbrand}, \emph{producttype}, \emph{maincategory}) by the Top-$k$ recommendation which $k$ is increasing from 1 to 10 and to calculate the recall rate and precise rate of them. In addition, we propose three models. In summary, we can get the 60 data points drawn in the Fig.\ref{1-2} and the Fig.\ref{2-3} to compare the performances between the different models.

\subsection{Experimental Results}
\begin{figure*}[!ht]
	\centering
	\includegraphics[width=0.85\linewidth]{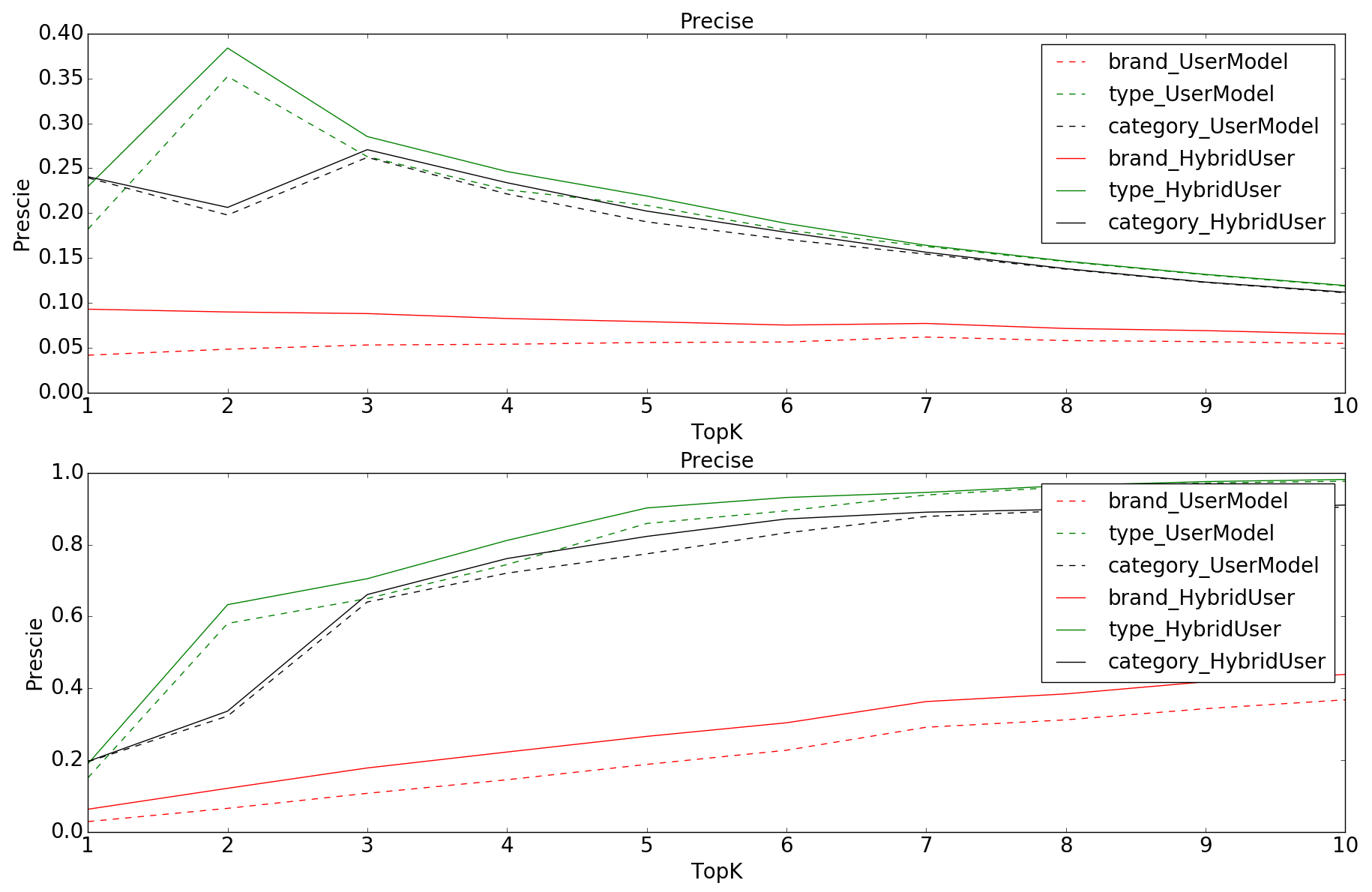}
	\caption{User Model and Hybrid User Model.}
	\label{1-2}
\end{figure*}

As Fig.\ref{1-2} shows, the dashed line represents the \emph{User Model} and the straight line represent the \emph{Hybrid User Model}. The recall rate curve of the \emph{User Model} and \emph{Hybrid User Model} both increase with $k$ increasing and The precise rate curve of the \emph{User Model} and \emph{Hybrid User Model} , when k is larger than 3, both decrease with $k$ increasing. The \emph{Hybrid User Model}, whether from accuracy or recall, is better than the \emph{User Model}.

\begin{figure*}[!ht]
	\centering
	\includegraphics[width=0.85\linewidth]{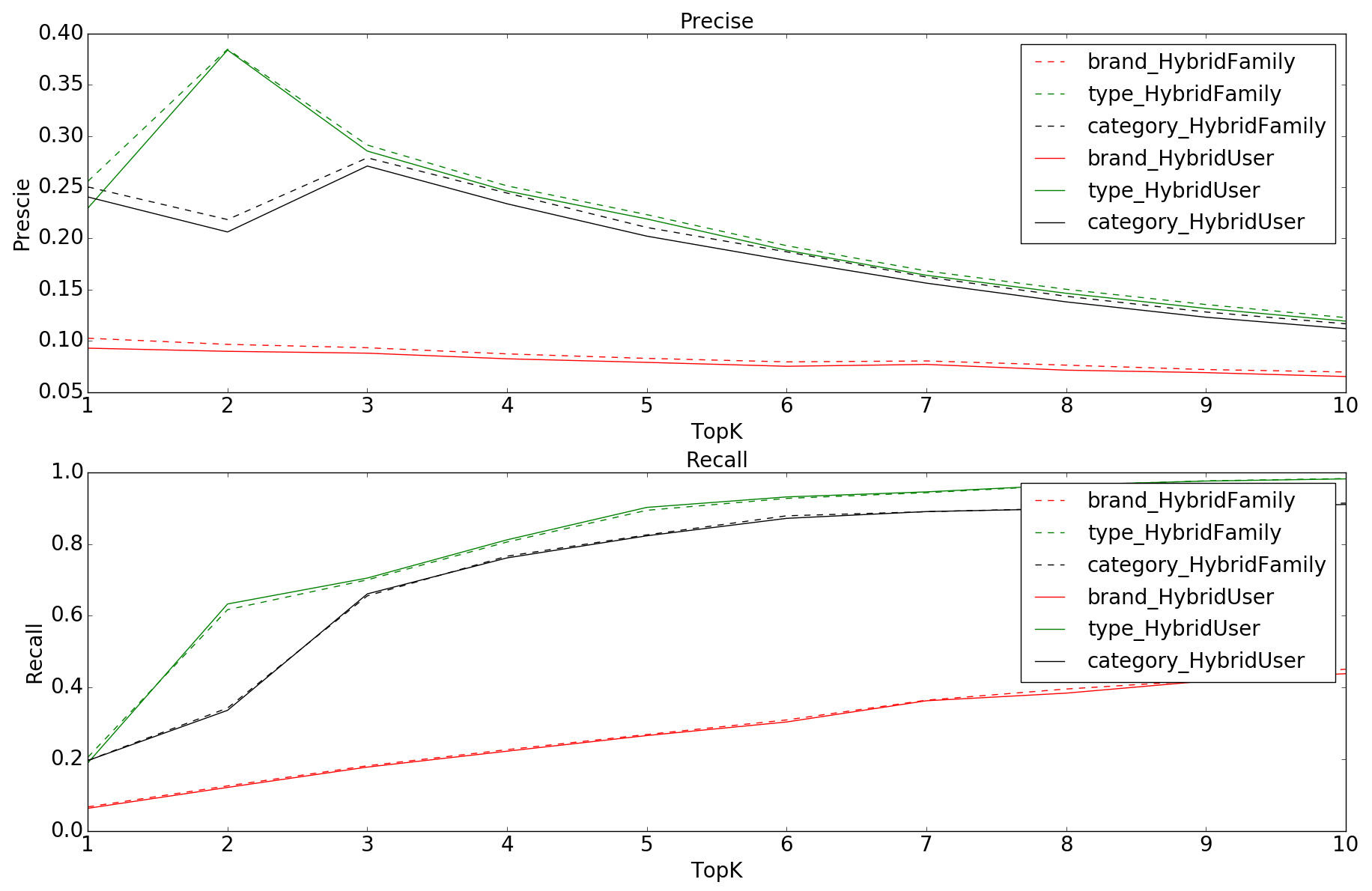}
	\caption{Hybrid User Model and Hybrid Family Model.}
	\label{2-3}
\end{figure*}

As Fig.\ref{2-3} shows, the straight line represents the \emph{Hybrid User Model} and the dashed line represent the \emph{Hybrid Family Model}. The recall rate curve of the \emph{Hybrid User Model} and the \emph{Hybrid User Model}  both increase with $k$ increasing and The precise rate curve of the \emph{Hybrid User Model} and the \emph{Hybrid Family Model}, when $k$ is larger than 3 both decrease with $k$ increasing. The \emph{Hybrid Family Model}, whether from accuracy or recall, is better than the \emph{Hybrid User Model}.

According to the conclusion above, the \emph{Hybrid Family Model}, for both accuracy and recall, is better than the \emph{User Model} and \emph{Hybrid User Model}.

\section{Conclusion and Future Work}\label{sec:con}
Firstly, we discuss the single user recommendation system and the popular group recommendation systems from multiple perspectives. After researching the dataset  , we propose a novel approach for integrating the information from user/family purchase data, activity data and shopping mall event data into an unified model for the family based recommendation problem. we verify our approach on via an empirical study on a real-world dataset and showcase the efficacy of our method. Finally, the algorithm works well according to the experimental results.

In the future, this algorithm can be combined with the matrix factorization or  other machine learning technology. This would be a good starting point for further research.






\bibliographystyle{IEEEtran}
\bibliography{IEEEabrv,papercite}
%

\end{document}